\documentclass[lettersize,journal]{IEEEtran}
\usepackage{amsmath,amsfonts}
\usepackage{algorithmic}
\usepackage{algorithm}
\usepackage{array}
\usepackage[caption=false,font=normalsize,labelfont=sf,textfont=sf]{subfig}
\usepackage{textcomp}
\usepackage{stfloats}
\usepackage{url}
\usepackage{verbatim}
\usepackage{graphicx}
\usepackage{cite}

\usepackage{bm}
\usepackage{amssymb}
\begin{document}

\title{Hybrid Beamforming and Waveform Design for Over-the-air Integrated Signal}

\author{Chonghao Zhao ~\IEEEmembership{}
       
\thanks{}
\thanks{}}

\markboth{Journal paper}
{Shell \MakeLowercase{\textit{et al.}}: A Sample Article Using IEEEtran.cls for IEEE Journals}

\maketitle

\begin{abstract}
The future wireless communications are expected to provide new use scenarios with emerging techniques. This paper focuses on vehicle to everything (V2X) network, where vehicles should cooperatively implement information obtaining, data sharing, and information postprocessing. Conventionally, the above three operations are considered in different layers or separated waveforms, leading to unavoidable interference and inefficient resource management when the number of devices becomes large. In this paper, we exploit the hybrid beamforming to design a cost-effective over-the-air integrated framework, and further consider the integrated waveform design problem with the constant-modulus constrain (CMC) and similarity constraint (SC). To solve these non-convex problems, an alternating optimization (AO) approach is proposed to jointly optimize the digital precoder as well as the hybrid combiner, using the successive convex approximation (SCA) and Riemannian conjugate gradient (RCG) algorithm. Additionally, we use the semidefinite relaxation (SDR) method to handle the practical waveform design problem. Numerical results demonstrate the effectiveness of the proposed hybrid beamforming and waveform design. 
\end{abstract}


\section{Introduction}
\IEEEPARstart{T}{o} support prosperous applications such as urban traffic, telemedicine, and agriculture in future society, the next-generation mobile communication is going to provide ubiquitous services to better protect the vulnerable  populations and advance human's well-being. In massive distributed networks, the limited energy, time, and bandwidth resources cannot satisfy the demand for high-accuracy and high-throughput communication. Furthermore, it is also impractical to share data among devices using any traditional orthogonal multi-access scheme, due to the mobility or excessive latency. Nowadays, integrated signals have finally made perception and communication move from separation to integration, bringing both integration gain and coordination gain. On the other hand, over-the-air data aggregation is another technology that can exploit the waveform superposition property to aggregate distributed data with limited radio resources and stringent latency \cite{zhu2021over}. To this end, the two emerging techniques can be naturally combined to facilitate both promising performances, while improving the spectrum efficiency and mitigating interference.

Several significant research has been conducted on integrated signal design. The signal processing techniques for joint design integrated systems have been largely analyzed. To optimize the transmit beampattern, optimization of the integrated beamforming in both coexistence and codesign criteria is analysed. The researchers later propose a joint beamforming model for integrated multiuser multiple-input multiple-output (MIMO) communication. Other work proposes a novel HBF-based massive MIMO integrated system based on the frameworks of the hybrid beamforming (HBF) communication and other relative system. Furthermore, the constant modulus integrated waveform was also considered. 

The pioneering work of over-the-air aggregation can be traced back to \cite{nazer2007computation}, and it is demonstrated in \cite{buck1976approximate} that any continuous function can be decomposed into a series of nomographic functions. In this sense, the antenna array is first introduced into it in \cite{chen2018over}, in which the intranode interference, nonuniform fading, and channel state information (CSI) were considered. To improve the aggregation mean squared error (MSE) in a cost-effective way, the work \cite{zhai2021hybrid} exploits the HBF to enhance over-the-air aggregation. Despite that, equalization and channel feedback scheme for MIMO over-the-air aggregation in mobility scenarios is also researched.

However, there are only a few preliminary works about the over-the-air integrated framework. Some considers the MSE to optimize beamforming design in both shared schemes and separated schemes. Later, other works investigated omnidirectional and directional beampattern designs, and further considered the balanced designs to coordinate the performance of three kinds of tasks. 

Generally, massive HBF \cite{yu2016alternating} is a promising solution to make a tradeoff between the equipment cost and array performance for MIMO communications. Moreover, the considered  constant-modulus constraint (CMC) makes the use of an efficient nonlinear power amplifier possible, and the similarity constraint (SC) leads to better range resolution and peak-to-average power ratio (PAPR). To the best of our knowledge, the HBF and waveform design of integrated signal is still an open issue.

In this paper, we focus on optimization-based signal designs for a MIMO integrated system. At first, we jointly optimize the digital precoder at the transmitters, and the fully-connected hybrid beamforming with phase shifters at the receiver to minimize the over-the-air aggregation MSE while guaranteeing other performance under power constraints. To address more practical CMC and SC, we further research the integrated waveform design with the linear frequency modulation signal as the benchmark signal. 

The remainder of this paper is organized as follows. Section II introduces the system model and formulates the optimization problems. In Section III, an HBF design problem is solved to balance the different performances. Section IV further considers the practical waveform design. Numerical results are provided in Section V, followed by the conclusion in Section VI.

Notations: Unless otherwise specified, we use bold upper-case letters (i.e., $\bm{H}$) for matrices, bold lower-case letters (i.e., $\bm{h}$) for vectors, and normal font for scalar (i.e., $\rho$). Subscripts (i.e., $i$, $j$) indicate the location of the entry in the matrices or vectors, and $\left(\cdot\right)^*$, $\left(\cdot\right)^T$, $\left(\cdot\right)^H$ and $\left(\cdot\right)^{\dagger}$ stand for complex conjugate, transpose, Hermitian transpose, and pseudo-inverse of a matrix, respectively. Furthermore, $tr\left(\cdot\right)$ and $vec\left(\cdot\right)$ denote the trace and the vectorization operations, $diag\left(\cdot\right)$ denotes the vector formed by the diagonal elements of the matrices, and $ddiag\left(\cdot\right)$ sets all off-diagonal elements as zero. $arg\left(\cdot\right)$ represents argument of a complex number, $\rm{Re}\left(\cdot\right)$ obtains the real part of the argument, and $\mathbb{E}\left[\cdot\right]$ represents the expectation. Moreover, $\Vert \cdot\Vert_2$, $\Vert \cdot\Vert_F$ are the $l_2$ norm and the Frobenius norm, while $\circ$, $\otimes$ and $<\cdot,\cdot>$ are the Hadamard product, Kronecker product and inner product, respectively.

\section{System Model and Problem Formulation}
The considered MIMO integrated system model has $K$ user equipments (UEs) transmit integrated waveforms to simultaneously obtain the information and aggregate data to the access point (AP). For simplicity, we assume that perfect synchronization among different UEs can be achieved using a reference clock \cite{abari2015airshare}.

\subsection{System Model}
In this model, each UE is equipped with a uniform linear array (ULA) of $N_t$ transmitting antennas, and the AP is equipped with $N_a$ antennas and $N_{rf}$ radio frequency (RF) chains for receiving. The data frame of UE $k$ is expressed as $\bm{S}_k=\left[\bm{s}_{k1},\bm{s}_{k2},...,\bm{s}_{kT}\right] \in \mathbb{C}^{M\times T}$, where $T$ presents the length of the frame and $M$ is the number of functions to be calculated. Without loss of generality, we assume that $M=N_t$ and data symbols $\bm{S}_k[t] \in \mathbb{C}^{M\times1}$ at arbitrary time slot $t$ are independent and identically distributed (i.i.d) among different UEs and functions with unit variance and zero mean, i.e., $\mathbb{E}[\bm{s}_k\bm{s}_k^H] = \bm{I}$, $\mathbb{E}[\bm{s}_k\bm{s}_j^H] = 0$, $\forall j\neq k$. 

At the time slot $t$, the aggregated data symbol vector received by the AP is given by
\begin{equation}\label{eq1}
\bm{y}_{AP}[t] = \bm{A}^H \sum_{k=1}^{K}\bm{H}_k\underbrace{\bm{F}_k\bm{S}_k[t]}_{\bm{X}_k[t]}+\bm{A}^H\bm{n}_c[t]
,\end{equation}
where $\bm{n}_c \in \mathbb{C}^{N_a\times1}$ is the additive white Gaussian noise (AWGN) vector with distribution $\mathcal{CN}(0,\sigma^2)$. Note that $\bm{A}^H = \bm{U}_{bb}^H\bm{U}_{rf}^H \in \mathbb{C}^{M\times N_a}$ is the hybrid aggregation combiner, where $\bm{U}_{bb}^H \in \mathbb{C}^{M\times N_{rf}}$ and $\bm{U}_{rf}^H \in \mathbb{C}^{N_{rf}\times N_a}$ are receive digital beamforming and analog beamforming matrix, respectively. In this case, it is assumed that $M\leq N_t$, $M\leq N_{rf}\leq N_a$, and $\lvert\bm{U}_{rf} (i,j)=1\rvert$, $\forall i\in N_r$, $\forall j\in N_{rf}$. For each UE $k$, $\bm{H}_k \in \mathbb{C}^{N_a\times N_t}$ is the channel matrix, and $\bm{X}_k[t] \in \mathbb{C}^{N_t\times1}$ is the $t$ th transmitted symbol of the transmitted integrated signal matrix $\bm{X}_k = \bm{F}_k\bm{S}_k \in  \mathbb{C}^{N_t\times T}$, in which $\bm{F}_k \in \mathbb{C}^{N_t\times M}$ is the digital precoder.

On the other hand, we closely match the desired beampattern and reference signal to guarantee the high peak side lobe ratio (PSLR) property and pulse compression performance, respectively. The beampattern can be expressed as 
\begin{equation}\label{eq2}
P_d(\theta) = \bm{\alpha}^H(\theta)\bm{X}_k[t]\bm{X}^H_k[t]\bm{\alpha}(\theta) = \bm{\alpha}^H(\theta)\bm{R}_{\bm{X}}\bm{\alpha}(\theta)
,\end{equation}
where $\bm{\alpha(\theta)}=\left[1,e^{j2\pi\Delta sin(\theta)},...,e^{j2\pi N_{t-1}\Delta sin(\theta)} \right] \in \mathbb{C}^{N_t\times 1}$ is the steering vector with $\Delta=d/\lambda$ being the ratio of the antenna spacing $d$ to the signal wavelength $\lambda$ and $\theta$ being the azimuth angle. Note that designing the beampattern $P_d$ is equivalent to optimizing the corresponding covariance matrix $\bm{R}_{\bm{X}}\in \mathbb{C}^{N_t\times N_t}$. At the $t$ th slot, the reference signal transmitted from antenna $n$ is 

\begin{equation}\label{eq3}
\bm{X}_0(n,t) = \sqrt{\frac{P}{N_t}} e^{j 2 \pi \left(f_0+(n-1)\Delta f\right)\left(t-1\right)}\cdot e^{j\pi k \left(t-1\right)^2}
,\end{equation}
where $k$ is the frequency modulation slope, $f_0$ and $\Delta f$ are the initial frequency and frequency spacing, respectively.

As the goal of over-the-air aggregation is to implement accurate multi-modal data fusion, its performance can be measured by the MSE of the $\bm{y}_{AP}$ distortion.
\begin{equation}\label{eq4}
\begin{aligned}
&{\rm MSE}({\bm{F}_k,\bm{U}_{rf},\bm{U}_{bb}})
=\mathbb{E}_t\left[\Vert\bm{y}_{AP}-\bm{y}\Vert_2^2\right]\\
&=\mathbb{E}_t\left[\Vert\sum_{k=1}^{K} \bm{A}^H\bm{H}_k\bm{X}_k[t]-\bm{S}_k[t]+\bm{A}^H\bm{n}_c\Vert_2^2\right]\\
&=\sum_{k=1}^{K}tr\left[\left(\bm{U}_{bb}^H\bm{U}_{rf}^H\bm{H}_k\bm{F}_k-\bm{I} \right)\left(\bm{U}_{bb}^H\bm{U}_{rf}^H\bm{H}_k\bm{F}_k-\bm{I} \right)^H\right]\\
&+\sigma^2tr\left(\bm{U}_{bb}^H\bm{U}_{rf}^H\bm{U}_{rf}\bm{U}_{bb}\right).
\end{aligned}
\end{equation}

\subsection{Problem Formulation}
Regarding the design of HBF, our goal is to minimize MSE of both the aggregation and the beampattern mismatch through the joint optimization of $\bm{F}_k$ at each UE, as well as $\bm{U}_{rf}$ and $\bm{U}_{bb}$ at the AP, under the per-antenna power constraint. The formulated weighted optimization problem is as follows,
\begin{align}\quad \min _{\bm{U}_{bb},\bm{U}_{rf},\{\bm{F}_k\}}\label{eq5}
&~~\rho\left(\sum_{k=1}^{K}\Vert \bm{U}_{bb}^H\bm{U}_{rf}^H\bm{H}_k\bm{F}_k-\bm{I}_{N_t}\Vert_F^2+\sigma^2\Vert \bm{A} \Vert_F^2\right)\nonumber\\ 
&~~+\left(1-\rho\right)\Vert\bm{F}_k-\bm{F}_0\Vert_F^2\\
\mathrm {s.t.}
&~diag\left(\bm{F}_k\bm{F}_k^H\right)=\frac{P}{N_t}\textbf{1}_{N_t},~ \forall k \in K , \tag{\ref{eq5}a}  \\
&~\lvert\bm{U}_{rf}(i,j)\rvert=1,~ \forall i \in N_r,~ \forall j \in N_{rf}  \tag{\ref{eq5}b}
,\end{align}
where $0\leq\rho\leq1$ is a weighting factor that determines the tradeoff between two kinds of performance and $\bm{F}_0$ is the beamforming matrix with strict ideal constraint.

The HBF only considers the beamforming matrices and thus does not guarantee to generate constant modulus waveforms. To achieve low modulus variation and high range resolution, here we consider the integrated waveform design that minimizes the aggregation MSE, subject to CMC and SC. The optimization problem can be given as 
\begin{align}\quad \min _{\{\bm{X}_k\}}  \label{eq6}
&~~\sum_{k=1}^{K}\Vert \bm{A}^H\bm{H}_k\bm{X}_k-\bm{S}_{k}\Vert_F^2+\sigma^2\Vert \bm{A} \Vert_F^2\\
\mathrm {s.t.}
&~\Vert\bm{X}_k-\bm{X}_0\Vert_F^2\leq \zeta,~ \forall k \in K ,  \tag{\ref{eq6}a} \\
&~\lvert\bm{X}_{k}(i,j)\rvert=\sqrt{\frac{P}{N_t}},~ \forall i \in N_t,~ \forall j \in T \tag{\ref{eq6}b}
,\end{align}
where $\zeta$ denotes the tolerable difference, and $\bm{X}_0$ is the reference waveform with constant-modulus.

The aforementioned two issues are non-convex, rendering them challenging to resolve. To achieve optimal solutions, we propose a series of algorithms, from Algorithm \ref{alg1} to Algorithm \ref{alg4} described in the following sections, that leverage the inherent structure of the problems.

\section{Hybrid Beamforming Design}
In this section, we propose an alternating optimization (AO) framework to iteratively update three highly coupled variables $\bm{F}_k$, $\bm{U}_{rf}$, and $\bm{U}_{bb}$ in integrated signal HBF design. In view of this, each subproblem becomes tractable and the whole algorithm can be finally converged.

\subsection{Transmit Digital Precoder $\bm{F}_k$ Design}
Initially, we fix the value of $\bm{U}_{rf}$ and $\bm{U}_{bb}$ to optimize $\bm{F}_k$ at each UE. In this case, problem (\ref{eq5}) can be equivalently decomposed into the following $K$ subproblems. 
\begin{align}\quad \min _{\bm{F}_k}  \label{eq7}
&~~\rho\left(\Vert \bm{A}^H\bm{H}_k\bm{F}_k-\bm{I}_{N_t}\Vert_F^2\right)+\left(1-\rho\right)\Vert\bm{F}_k-\bm{F}_0\Vert_F^2\\
\mathrm {s.t.}
&~diag\left(\bm{F}_k\bm{F}_k^H\right)=\frac{P}{N_t}\textbf{1}_{N_t} \tag{\ref{eq7}a}  
.\end{align}
Note that the benchmark beamformer $\bm{F}_0 \in \mathbb{C}^{N_t\times M}$ can be obtained through the Orthogonal Procrustes problem (OPP) in \cite{gower2004procrustes}, which is given as
\begin{equation}\label{eq8}
\bm{F}_0=\bm{Q}_k\bm{U}_k\bm{I}_{N_t\times M}\bm{V}_k^H
,\end{equation}
where the desired covariance matrix $\bm{R}_k$ is decomposed by $\bm{R}_k=\bm{Q}_k\bm{Q}_k^H$, and $\bm{Q}_k^H\bm{H}_k^H\bm{A}=\bm{U}_k\bm{\Sigma}_k\bm{V}_k^H$ is the SVD.

By the definition of the Frobeniushe norm, problem (\ref{eq7}) can be written more compactly as 
\begin{align}\quad \min _{\bm{F}_k}  \label{eq9}
&~~\Vert\bm{C}\bm{F}_k-\bm{D}\Vert_F^2\\
\mathrm {s.t.}
&~diag\left(\bm{F}_k\bm{F}_k^H\right)=\frac{P}{N_t}\textbf{1}_{N_t} \tag{\ref{eq9}a}
,\end{align}
where $\bm{C}=\left[\sqrt{\rho}\left(\bm{A}^H\bm{H}_k\right)^T,\sqrt{1-\rho}\bm{I}_{N_t}\right]^T \in \mathbb{C}^{\left(M+N_t\right)\times N_t}$, $\bm{D}=\left[\sqrt{\rho}\bm{I}_{M},\sqrt{1-\rho}\bm{F}_{0}^T\right]^T\in \mathbb{C}^{\left(M+N_t\right)\times M}$, and $\textbf{1} = \left[1, 1, . . . , 1\right]^T \in \mathbb{R}^{N_t\times 1}$ is the all-one vector. The reformulated problem is still non-convex quadratic constrained quadratic programming (QCQP) since the diagonal constraint can be regarded as $N$ quadratic equality constraints. 

To deal with problem (\ref{eq9}), we propose a Riemannian conjugate gradient (RCG) algorithm that can achieve a near-optimal solution within relatively low complexity. It is worth noting that the feasible set $\mathcal{M}$ of the problem is a complex oblique manifold. Therefore, we can solve the rewritten unconstrained least squares (LS) problem on a manifold, which is 
\begin{equation}\label{eq10}
\quad \min _{\bm{F}_k\in\mathcal{M} } 
~~\Vert \bm{C}\bm{F}_k-\bm{D} \Vert_F^2.
\end{equation}
We denote the $i$ and $\bm{F}_k^i$ as the steps and the point to be updated at step $i$, respectively. On this basis, the tangent space $T_{\bm{F}_k^i}\mathcal{M}$ for the manifold $\mathcal{M}$ at a given point $\bm{F}_k^i$ is defined as 
\begin{equation}\label{eq11}
T_{\bm{F}_k^i}\mathcal{M}=\left\{\bm{Z}\in \mathbb{C}^{N_t\times M}|{\rm Re}\left(\left[\bm{F}_k^{iH}\bm{Z}\right]_{nn}\right)=0,~ \forall n \right\}
,\end{equation}
and the objective function is $f\left(\bm{F}_k\right)$ . The Euclidean gradient is thus given by 
\begin{equation}\label{eq12}
\nabla_{\bm{F}_k}f=2\bm{C}^H\left(\bm{C}\bm{F}_k-\bm{D}\right)
,\end{equation}
and the Riemannian gradient can be calculated by projecting (\ref{eq12}) onto $T_{\bm{F}_k^i}\mathcal{M}$,
\begin{equation}\label{eq13}
grad_{\bm{F}_k^i} f=\nabla_{\bm{F}_k}f-\bm{F}_k^{iH} ddiag\left({\rm Re}\left(\nabla_{\bm{F}_k}f\right)^H\bm{F}_k^i\right)
.\end{equation}
Similar to the conjugate gradient (CG) algorithm, in the RCG method the stepsize $\delta_i$ is obtained by the Armijo rule, and the descent direction $\bm{\mu}_i$ is updated as a nonlinear combination of the $grad_{\bm{F}_k^i} f$ of the current iteration $i$ and the descent direction of the previous iteration $i-1$,
\begin{equation}\label{eq14}
\bm{\mu}_i=-grad_{\bm{F}_k^{i}}f+\alpha_{i-1}\mathcal{T}_{\bm{F}_k^{i-1}\rightarrow\bm{F}_k^{i}}\left(\bm{\mu}_{i-1}\right)
,\end{equation}
where the combination coefficient $\alpha_{i-1}$ is calculated by the Polak-Ribi\'{e}re formula. It is worth noting that $\mathcal{T}_{\bm{F}_k^{i-1}\rightarrow\bm{F}_k^{i}}\left(\bm{\mu}_{i-1}\right)$ denotes the transportation process that project $\bm{\mu}_{i-1}$ from $T_{\bm{F}_k^{i-1}}\mathcal{M}$ to $T_{\bm{F}_k^i}\mathcal{M}$. To finally map the point $\bm{Z}$ on $T_{\bm{F}_k^i}\mathcal{M}$ back to $\mathcal{M}$, we then define a specific mapping called retraction, which can be written as
\begin{equation}\label{eq15}
\begin{aligned}
\bm{F}_k^{i+1}&=\mathcal{R}_{\bm{F}_k^i}\left(\delta_i\bm{\mu}_{i}\right) \triangleq T_{\bm{F}_k^i}\mathcal{M}\mapsto \mathcal{M}\\
&=\sqrt{\frac{P}{N_t}} ddiag\left(\left(\bm{F}_k^i+\bm{Z}\right)\left(\bm{F}_k^i+\bm{Z}\right)^H\right)^{-1/2}\left(\bm{F}_k^i+\bm{Z}\right).
\end{aligned}
\end{equation}

The main procedure of RCG algorithm is summarized in Algorithm \ref{alg1}.
\begin{algorithm}[H]
\caption{RCG Algorithm for Solving Problem (\ref{eq7}).}\label{alg1}
\begin{algorithmic}
\STATE 
\STATE  \textbf{Input: } $\bm{H}_k$, $\bm{A}$, $\bm{F}_0$, $P$, $0\leq\rho\leq1$, $\epsilon>0$, $I_{max}>2$, $i=1$
\STATE \textbf{Output: } $\bm{F}_k$ 
\STATE \textbf{Initialization, }obtain $\bm{C}$, $\bm{D}$.
\STATE \textbf{While }$i\leq I_{max}$ and $\Vert grad_{\bm{F}_k^i} f\Vert_F \geq \epsilon$, \textbf{do }
\STATE \hspace{0.5cm}1. Compute $\alpha_{i-1}$ using Polak-Ribi\'{e}re formula as
\STATE \hspace{1.0cm}$\alpha_{i-1}=\frac{\langle~ grad_{\bm{F}_k^{i}}f~,~grad_{\bm{F}_k^{i}}f-\mathcal{T}_{\bm{F}_k^{i-1}\rightarrow\bm{F}_k^{i}}\left(grad_{\bm{F}_k^{i-1}}f\right)\rangle}{~\langle grad_{\bm{F}_k^{i-1}}f~ ,~grad_{\bm{F}_k^{i-1}}f~\rangle}$;
\STATE \hspace{0.5cm}2. Calculate the descent direction $\bm{\mu}_i$ by
\STATE \hspace{1.0cm} $\bm{\mu}_i=-grad_{\bm{F}_k^{i}}f+\alpha_{i-1}\mathcal{T}_{\bm{F}_{i-1}\rightarrow\bm{F}_{i}}\left(\bm{\mu}_{i-1}\right)$;
\STATE \hspace{0.5cm}3. Calculate the stepsize $\delta_i$ by Armijo search method;
\STATE \hspace{0.5cm}4. Update $\bm{F}_k^{i+1}$ by $\bm{F}_k^{i+1}=\mathcal{R}_{\bm{F}_k^i}\left(\delta_i\bm{\mu}_{i}\right)$;
\STATE \hspace{0.5cm}5. $i=i+1$;
\STATE \textbf{end While}.
\end{algorithmic}
\end{algorithm}

\subsection{Receive Analog Beamforming $\bm{U}_{rf}$ Design}
Given the ${\bm{F}_k}$ and $\bm{U}_{bb}$, in this subsection, we optimize the $\bm{U}_{rf}$. The corresponding problem formulation can be given as
\begin{align} \quad \min _{\bm{U}_{rf},\bm{U}_{rf}}  \label{eq16}
&~~tr\left(\left(\bm{U}_{bb}^H\bm{U}_{rf}^H\bm{H}_k\bm{F}_k-\bm{I} \right)\left(\bm{U}_{bb}^H\bm{U}_{rf}^H\bm{H}_k\bm{F}_k-\bm{I} \right)^H\right) \nonumber \\
&~~+\sigma^2tr\left(\bm{U}_{bb}^H\bm{U}_{rf}^H\bm{U}_{rf}\bm{U}_{bb}\right)\\
\mathrm {s.t.}
&~\lvert\bm{U}_{rf}(i,j)\rvert=1,~ \forall i \in N_r,~ \forall j \in N_{rf}  \tag{\ref{eq16}a}
,\end{align}
which is also non-convex due to the constant constraints.

To tackle this issue, we recast the constraint in its exponential form, i.e., $\bm{U}_{rf}(i,j)=e^{\sqrt{-1}\bm{\theta}\left(\left(j-1\right)N_r+i\right)},~\bm{\theta}=\left[\theta_1,\theta_2,...,\theta_{N_rN_{rf}}\right]^T,~ \forall i \in N_r,~ \forall j \in N_{rf}$, and thus the problem can be further transformed as
\begin{align} \quad \min _{\bm{\theta}}  \label{eq17}
&~~tr\left(\left(\bm{U}_{bb}^H\bm{U}_{rf}^H\left(\bm{\theta}\right)\bm{H}_k\bm{F}_k-\bm{I} \right)\left(\bm{U}_{bb}^H\bm{U}_{rf}^H\left(\bm{\theta}\right)\bm{H}_k\bm{F}_k-\bm{I} \right)^H\right) \nonumber \\
&~~+\sigma^2tr\left(\bm{U}_{bb}^H\bm{U}_{rf}^H\left(\bm{\theta}\right)\bm{U}_{rf}\left(\bm{\theta}\right)\bm{U}_{bb}\right)\\
\mathrm {s.t.}
&~-\pi\leq\bm{\theta}(i)\leq\pi,~ \forall i \in \bm{\psi},~ \bm{\psi}=\left[1,2,...,N_rN_{rf}\right] \tag{\ref{eq17}a}
.\end{align}
In this way, the intractable constraints are rewritten as linear constraints, while the objective function becomes non-convex. To this end, we propose a successive convex approximation (SCA) algorithm in which a surrogate function $\hat{f}(\bm{\theta},\bm{\theta}_r)$ of the objective function $f(\bm{\theta})$ is optimized and then used to approximate the original function.

Specifically, we use the classic first-order Taylor approximation to develop the surrogate function which is written as
\begin{equation}\label{eq18}
\hat{f}\left(\bm{\theta},\bm{\theta}_r\right)=f\left(\bm{\theta}_r\right)+\gamma_{\bm{\theta}_r}^H\left(\bm{\theta}-\bm{\theta}_r\right)+\tau\Vert\bm{\theta}-\bm{\theta}_r\vert_2^2
,\end{equation}
where $r$ and $\bm{\theta}_r$ are the iteration number and the $r$ th iteration point, respectively. Here, $\tau$ is a small positive number to guarantee the convexity and to adjust the convergence rate. The gradient of $\hat{f}(\bm{\theta},\bm{\theta}_r)$ at point $r$ is given by
\begin{equation}\label{eq19}
\gamma_{\bm{\theta}_r}=\nabla_{\bm{\theta}_r} f\left(\bm{\theta}\right)\big|_{\bm{\theta} = \bm{\theta}_r} =-vec\left\{2{\rm Re}\left[\sqrt{-1}\bm{U}_{rf,r}^{*}\circ\bm{F}_r\right]\right\}
,\end{equation}
in which  
\begin{equation}\label{eq20}
\begin{aligned}
\bm{F}_r&=\left(\sum_{k=1}^K\bm{H}_k\bm{F}_k\bm{F}_k^H\bm{H}_k^H+\sigma^2\bm{I}\right)\bm{U}_{rf,r}\bm{U}_{bb}\bm{U}_{bb}^H\\
&-\sum_{k=1}^K\bm{H}_k\bm{F}_k\bm{U}_{bb}^H
\end{aligned}
.\end{equation}

It can be observed that $\hat{f}(\bm{\theta},\bm{\theta}_r)$ is a continuous function and it has the same values and gradient as the $f(\bm{\theta})$ at point $r$. Hence, it satisfies the requirements of the surrogate function. The problem in the SCA algorithm is finally reformulated as 
\begin{align} \quad \min _{\bm{\theta}} \label{eq21} 
&~~\hat{f}\left(\bm{\theta},\bm{\theta}_r\right)\\
\mathrm {s.t.}
&~-\pi\leq\bm{\theta}(i)\leq\pi,~ \forall i \in \bm{\psi} \tag{\ref{eq21}a}
.\end{align}
The proposed SCA method is provided in Algorithm \ref{alg2}.
\begin{algorithm}[H]
\caption{SCA Method for Solving Problem (\ref{eq16}).}\label{alg2}
\begin{algorithmic}
\STATE 
\STATE  \textbf{Input: } $\bm{H}_k$, $\bm{U}_{bb}$, $\bm{F}_k$, $\tau>0$, $\epsilon>0$, $r=1$ 
\STATE \textbf{Output: } $\bm{U}_{rf}$ 
\STATE \textbf{Initialization. }
\STATE \textbf{While }$\lvert f\left(\bm{\theta}\right)\big|_{\bm{\theta} = \bm{\theta}_r}-f\left(\bm{\theta}\right)\big|_{\bm{\theta} = \bm{\theta}_{r-1}}\rvert>\epsilon$, \textbf{do }
\STATE \hspace{0.5cm}1. Compute the gradient $\gamma_{\bm{\theta}_r}$ by 
\STATE \hspace{1.0cm}$\gamma_{\bm{\theta}_r}=-vec\left\{2{\rm Re}\left[\sqrt{-1}\bm{U}_{rf,r}^{*}\circ\bm{F}_r\right]\right\}$;
\STATE \hspace{0.5cm}2. Update $\bm{\theta}_{r+1}$ by
\STATE \hspace{1.0cm}$\bm{\theta}_{r+1}\left(i\right)=mod\left(\bm{\theta}_r\left(i\right)-\frac{\gamma_{\bm{\theta}_r}\left(i\right)}{2\tau},2\pi\right),~ \forall i \in \bm{\psi}$; 
\STATE \hspace{0.5cm}3. Update $\bm{U}_{rf,r+1}$ by 
\STATE \hspace{1.0cm}$\bm{U}_{rf,r+1}\left(i,j\right)=e^{\sqrt{-1}\bm{\theta}_{r+1}\left(\left(j-1\right)N_r+i\right)},~ \forall i \in N_r,~ \forall j \in N_{rf}$; 
\STATE \hspace{0.5cm}4. $r=r+1$;
\STATE \textbf{end While}.
\end{algorithmic}
\end{algorithm}

\subsection{Receive Digital Beamforming $\bm{U}_{bb}$ Design}
At last, the $\bm{U}_{bb}$ can be optimized under given ${\bm{F}_k}$ and $\bm{U}_{rf}$, which is an unconstrained convex problem.
\begin{equation}\label{eq22}
\begin{aligned}
\quad \min _{\bm{U}_{bb}} 
&~~tr\left(\left(\bm{U}_{bb}^H\bm{U}_{rf}^H\bm{H}_k\bm{F}_k-\bm{I} \right)\left(\bm{U}_{bb}^H\bm{U}_{rf}^H\bm{H}_k\bm{F}_k-\bm{I} \right)^H\right) \\
&~~+\sigma^2tr\left(\bm{U}_{bb}^H\bm{U}_{rf}^H\bm{U}_{rf}\bm{U}_{bb}\right).
\end{aligned}
\end{equation}
The updated $\bm{U}_{bb}$ can then be given by
\begin{equation}\label{eq23}
\begin{aligned}
\bm{U}_{bb}&=\left(\bm{U}_{rf}^H\left(\sum_{k=1}^{K}\bm{H}_k\bm{F}_k\bm{F}_k^H\bm{H}_k^H+\sigma^2\bm{I}\right)\bm{U}_{rf}\right)^{-1}\\
&\times\bm{U}_{rf}^H\left(\sum_{k=1}^{K}\bm{H}_k\bm{F}_k\right)
\end{aligned}
.\end{equation}

\subsection{Complexity Analysis and Overall Algorithm}
We summarize this section by analyzing the whole algorithm. For the computational complexity, we only consider the dominant part and omit the low-order terms. Since the number of iterations is unpredictable, we only analysis it in each iteration and the convergence performance will be further investigated in section V.

In Algorithm 1, the calculation of $\bm{F}_0$ involves Cholesky decomposition, matrix multiplications, and SVD, leading to a complexity of $\mathcal{O}(N_t^3+N_t^2M+N_tM^2+N_t^2N_a+N_tN_aM)$. Moreover, in the RCG algorithm the highest-order terms come from the Euclidean gradient (\ref{eq12}) and the Riemannian gradient (\ref{eq13}) calculation, i.e., $\mathcal{O}(M^2N_t+N_t^2M)$. The costs of obtaining transportation, retraction (\ref{eq15}), and inner product are $\mathcal{O}(N_t^2M)$. For Algorithm \ref{alg2}, the most costly step is the processing of the gradient in (\ref{eq19}), requiring complexity $\mathcal{O}(KN_aN_tM+KN_aMN_{rf}+KN_a^2N_t+KN_a^2N_{rf})$. At last, the complexity of calculating $\bm{U}_{bb}$ is $\mathcal{O}(KN_aN_tM+KN_{rf}N_aM+KN_a^2N_t+KN_a^2N_{rf}+KN_{rf}^2N_a)$. Hence, at each iteration, the overall algorithm complexity at the AP and each UE are $\mathcal{O}(KN_aN_tM+KN_{rf}N_aM+KN_a^2N_t+KN_a^2N_{rf}+KN_{rf}^2N_a)$ and $\mathcal{O}(N_t^3+N_t^2M+N_tM^2+N_t^2N_a+N_tN_aM)$, respectively.

Based on the aforementioned results, we can conclude that the majority of the processing are offloaded to the AP, while the algorithms on the UEs have relatively smaller complexity. To obtain the information required for the above processing, it is found that each UE only needs some general information such as $\bm{H}_k$, $\bm{U}_{rf}$, and $\bm{U}_{bb}$, which means it can easily acquire them through either the classic estimation method or AP broadcast, and do not need the knowledge of other UEs in the network. Additionally, AP needs the $\bm{F}_k$ and $\bm{H}_k$ of each UE. To handle this, we note that the related expressions in our algorithm, like $\sum{\bm{H}_k\bm{F}_k\bm{F}_k}^H\bm{H}_k^H$, all have the form of the nomographic functions. Therefore, the channel estimation and feedback can be unified and also integrated into this integrated framework, resulting in reduced latency for mobility applications. 

For clarity, we summarize the pseudo-code of the whole framework in Algorithm \ref{alg3}.
\begin{algorithm}[H]
\caption{AO framework for Solving Problem (\ref{eq5}).}\label{alg3}
\begin{algorithmic}
\STATE 

\STATE  \textbf{Input: } $\bm{H}_k$, $\{\bm{F}_k\}$, $\bm{U}_{rf}$, $\bm{U}_{bb}$, $N_{iteration}>2$, $n=1$
\STATE \textbf{Output: } $\{\bm{F}_k\}$, $\bm{U}_{rf}$, $\bm{U}_{bb}$ 
\STATE \textbf{Initialization. }
\STATE \textbf{While }$n<N_{iteration}$, \textbf{do }
\STATE \hspace{0.5cm}1. Optimize $\bm{F}_k,~ \forall k \in K$ according to Algorithm \ref{alg1} 
\STATE \hspace{0.5cm}2. Optimize $\bm{U}_{rf}$ according to Algorithm \ref{alg2};
\STATE \hspace{0.5cm}3. Optimize $\bm{U}_{bb}$ by
\STATE \hspace{1.0cm}$\bm{U}_{bb}=\left(\bm{U}_{rf}^H\left(\sum_{k=1}^{K}\bm{H}_k\bm{F}_k\bm{F}_k^H\bm{H}_k^H+\sigma^2\bm{I}\right)\bm{U}_{rf}\right)^{-1}\times\bm{U}_{rf}^H\left(\sum_{k=1}^{K}\bm{H}_k\bm{F}_k\right)$; 
\STATE \hspace{0.5cm}4. $n=n+1$;
\STATE \textbf{end While}.
\end{algorithmic}
\end{algorithm}

\section{Constant Modulus Waveform Design}
Based on the above consideration, we further consider more practical waveform-level design for integrated signal. The CMC and SC we introduced in this problem can empower it with good ambiguity characteristics and low PAPR.

Following the same analysis in the previous section, we divided problem (\ref{eq6}) into $K$ subproblems and each can be simplified as
\begin{align}\quad \min _{\bm{X}_k} \label{eq24}
&~~\Vert \bm{A}^H\bm{H}_k\bm{X}_k-\bm{S}_{k}\Vert_F^2\\
\mathrm {s.t.}
&~\Vert\bm{X}_k-\bm{X}_0\Vert_F^2\leq \zeta,   \tag{\ref{eq24}a} \\
&~\lvert\bm{X}_{k}(i,j)\rvert=\sqrt{\frac{P}{N_t}},~ \forall i \in N_t,~ \forall j \in T \tag{\ref{eq24}b}
,\end{align}
which is a non-convex QCQP. In this section, we first transform (\ref{eq24}) into a homogeneous form suitable for semidefinite relaxation (SDR) and then solve the semi-definite programming (SDP) problem resulting from relaxation. 

To be specific, the objective function can be vectorized as
\begin{equation}\label{eq25}
\begin{aligned}
&~~~~\Vert{\rm vec}\left(\bm{A}^H\bm{H}_k\bm{X}_k-\bm{S}_k\right)\Vert_2^2\\
&=\Vert{\rm vec}\left(\bm{A}^H\bm{H}_k\bm{X}_k\right)-{\rm vec}\left(\bm{S}_k\right)\Vert_2^2\\
&=\Vert{\rm vec}\left(\bm{S}_k\right)-\left(\bm{I}_T\otimes \bm{A}^H\bm{H}_k\right){\rm vec}\left(\bm{X}_k\right)\Vert_2^2
\end{aligned}
.\end{equation}
Here we define $\bm{x}_0={\rm vec}\left(\bm{X}_0\right)$, $\bm{b}={\rm vec}\left(\bm{X}_k\right)$, $\bm{s}={\rm vec}\left(\bm{S}_k\right)$, $\bm{E}=\bm{I}_T\otimes \bm{A}^H\bm{H}_k\cdot \sqrt{\frac{P}{N_t}}$ to further reformulate the problem as
\begin{align}\quad \min _{\bm{b}} \label{eq26}
&~~\Vert \bm{E}\bm{b}-\bm{s}\Vert_2^2\\
\mathrm {s.t.}
&~\Vert\bm{b}-\bm{x}_0\Vert_2^2\leq \zeta,   \tag{\ref{eq26}a} \\
&~\lvert\bm{b}_n\rvert=1,~ \forall n  \tag{\ref{eq26}b}
.\end{align}
Note that the objective function and (\ref{eq26}a) are nonhomogeneous, we thus introduce an auxiliary variable $t=1$ to convert them to quadratic form, which is given by
\begin{equation}\label{eq27}
\begin{aligned}
\Vert\bm{E}\bm{b}-\bm{s}\Vert_2^2&=\left(\bm{E}\bm{b}-\bm{s}\right)^H\left(\bm{E}\bm{b}-\bm{s}\right)\\
&=\bm{b}^H\bm{E}^H\bm{E}\bm{b}-\bm{s}^H\bm{E}\bm{b}-\bm{b}^H\bm{E}^H\bm{s}+\bm{s}^H\bm{s}\\
&=\begin{bmatrix}\bm{b}^H&t\end{bmatrix}\begin{bmatrix}\bm{E}^H\bm{E}&-\bm{E}^H\bm{s}\\-\bm{s}^H\bm{E}& \bm{s}^H\bm{s}\end{bmatrix} \begin{bmatrix}\bm{b}\\t\end{bmatrix}
\end{aligned}
,\end{equation}
\begin{equation}\label{eq28}
\begin{aligned}
\Vert\bm{b}-\bm{x}_0\Vert_2^2&=\left(\bm{b}-\bm{x}_0\right)^H\left(\bm{b}-\bm{x}_0\right)\\
&=\bm{b}^H\bm{b}-\bm{x}_0^H\bm{b}-\bm{b}^H\bm{x}_0+\bm{x}_0^H\bm{x}\\
&=\begin{bmatrix}\bm{b}^H&t\end{bmatrix}\begin{bmatrix}\bm{I}_{N_t T}&-\bm{x}_0\\-\bm{x}_0^H& \bm{x}_0^H\bm{x}_0\end{bmatrix} \begin{bmatrix}\bm{b}\\t\end{bmatrix}
\end{aligned}
.\end{equation}

In this case, the problem (\ref{eq26}) becomes the following binary quadratic programming (BQP),
\begin{align}\quad \min _{\bm{x}} \label{eq29}
&~~\bm{x}^H\bm{C}\bm{x}\\
\mathrm {s.t.}
&~\bm{x}^H\bm{D}\bm{x}\leq \zeta,   \tag{\ref{eq29}a} \\
&~\lvert\bm{x}_n\rvert=1,~ \forall n  \tag{\ref{eq29}b}
,\end{align}
where $\bm{C}=\begin{bmatrix}\bm{E}^H\bm{E}&-\bm{E}^H\bm{s}\\-\bm{s}^H\bm{E}& \bm{s}^H\bm{s}\end{bmatrix}$, $\bm{D}=\begin{bmatrix}\bm{I}_{N_t T}&-\bm{x}_0\\-\bm{x}_0^H& \bm{x}_0^H\bm{x}_0\end{bmatrix}$, and $\bm{x}=\begin{bmatrix}\bm{b}\\t\end{bmatrix}$. Moreover, by introducing a new variable $\bm{X}=\bm{x}\bm{x}^H$, the problem can be rewritten as
\begin{align}\quad \min _{\bm{X}}  \label{eq30}
&~~\bm{C}\bm{X}\\
\mathrm {s.t.}
&~\bm{D}\bm{X}\leq \zeta,   \tag{\ref{eq30}a} \\
&~\lvert\bm{X}_{nn}\rvert=1,~ \forall n,  \tag{\ref{eq30}b}\\
&~\bm{X}\succeq 0 \tag{\ref{eq30}c},\\
&~{\rm Rank}\left(\bm{X}\right)=1  \tag{\ref{eq30}d}
,\end{align}
which can be relaxed by applying the SDR algorithm. By omitting the rank-1 constraints, problem (\ref{eq30}) becomes the following classic SDP and a suboptimal solution can thus be efficiently obtained,
\begin{align}\quad \min _{\bm{X}}  \label{eq31}
&~~\bm{C}\bm{X}\\
\mathrm {s.t.}
&~\bm{D}\bm{X}\leq \zeta,   \tag{\ref{eq31}a} \\
&~\lvert\bm{X}_{nn}\rvert=1,~ \forall n,  \tag{\ref{eq31}b}\\
&~\bm{X}\succeq 0 \tag{\ref{eq31}c}
.\end{align}

The optimal solution $\bm{X}^*$ of the problem (\ref{eq31}) can be obtained using a convex solver and then we exploit the Gaussian randomization sampling to extract a set of optimized $\{\bm{\hat{x}}\}$ from $\bm{X}^*$. To retrieve vector $\bm{b}$, the introduced auxiliary variable $t=1$ in each $\bm{\hat{x}}$ can be eliminated by
\begin{equation}\label{eq32}
\bm{\hat{b}}=e^{j{\rm arg}\left(\left[\frac{\bm{\hat{x}}}{\bm{\hat{x}}_{N_tT+1}}\right]_{\left[1:N_tT\right]}\right)}
,\end{equation}
where $[\cdot]_{(1:N)}$ obtains the first $N$ elements in the original vector. At last, the optimized $\bm{X}_k^*$ is matrixed from the one among $\{\bm{\hat{b}}\}$ that achieves the minimum objective function and meets the constraints in (\ref{eq24}).

Note that the SDR approach does not have closed-form complexity expressions. The developed SDR method is shown as algorithm \ref{alg4}.
\begin{algorithm}[H]
\caption{SDR Method for Solving Problem (\ref{eq24}).}\label{alg4}
\begin{algorithmic}
\STATE 
\STATE  \textbf{Input: }$\bm{H}$, $\bm{A}$, $\bm{X}_0$, $\bm{S}_k$, $N>0$
\STATE \textbf{Output: } $\bm{X}_k$
\STATE \textbf{Initialization}, vectorize the matrix as $\bm{x}_0$, $\bm{b}$, $\bm{s}$, and obtain $\bm{E}$, $\bm{C}$, $\bm{D}$.
\STATE 1. Introduce a new varible $\bm{X}=\bm{x}\bm{x}^H$ in (\ref{eq30}) to reformulate the problem as SDP. 
\STATE 2. Perform eigen decomposition as $\bm{X}^* = \bm{U}\bm{\Sigma}\bm{U}^H$. 
\STATE 3. Retrieve $N$ feasible solutions using Gaussian random sampling $\bm{\hat{x}}=\bm{U}\bm{\Sigma}^{1/2}\bm{z}$, $\bm{z}\sim\mathcal{CN}(0,I_{N_tT+1})$.
\STATE 4. Eliminate the auxiliary variable $t$ by  
\STATE \hspace{0.5cm}$\bm{\hat{b}}=e^{j{\rm arg}\left(\left[\frac{\bm{\hat{x}}}{\bm{\hat{x}}_{N_tT+1}}\right]_{\left[1:N_tT\right]}\right)}$.
\STATE 5. Recover $\bm{X}_k^*$ from $\{\bm{\hat{b}}\}$.

\end{algorithmic}
\end{algorithm}

\section{Numerical Results}
In this section, we evaluate the effectiveness of the proposed HBF and waveform design algorithm, where the mismatch, convergence performance, and Pareto bounds are considered. The performance is evaluated through beampattern and waveform mismatch while the normalized aggregation MSE, i.e., ${\rm MSE}/M$, acts as the metric of the over-the-air aggregation performance. Unless otherwise stated, in all simulations we consider a network with integrated signal comprising one AP with $N_a=64$ antennas connected to $N_{rf}=N_a/4$ RF chains, and $K=5$ UEs, each with $M=N_t=16$. To simplify, here we set the maximum transmission power $P=1$, ${\rm SNR}=4$ dB, and all the channels are assumed to be i.i.d. Rayleigh fading with complex Graussian distribution $\mathcal{CN}(0,1)$. Moreover, the signal frequency is set as $f_c=5.9$ GHz, and $d$ is half of the wavelength. We set $\tau$ in SCA as 1.2, and the weighting factor is $\rho=0.03$. In HBF design, we assume the four primary directions are $[0.22\pi,0.39\pi,0.61\pi,0.78\pi]$ with the main lobe being $11^{\circ}$. As for waveform design, a reference signal with bandwidth $B=1$ MHz and pulse width $T_p= 0.0125$ ms is selected as the benchmark and $\zeta$ is set as $20$. In particular, we set $T$ as $50$, and the points are sampled with rates $f_s=4$ MHz.

\subsection{Resultant Integrated Beampattern and Waveform}

\begin{figure}[htbp]
\centerline{\includegraphics [width=3.0 in]{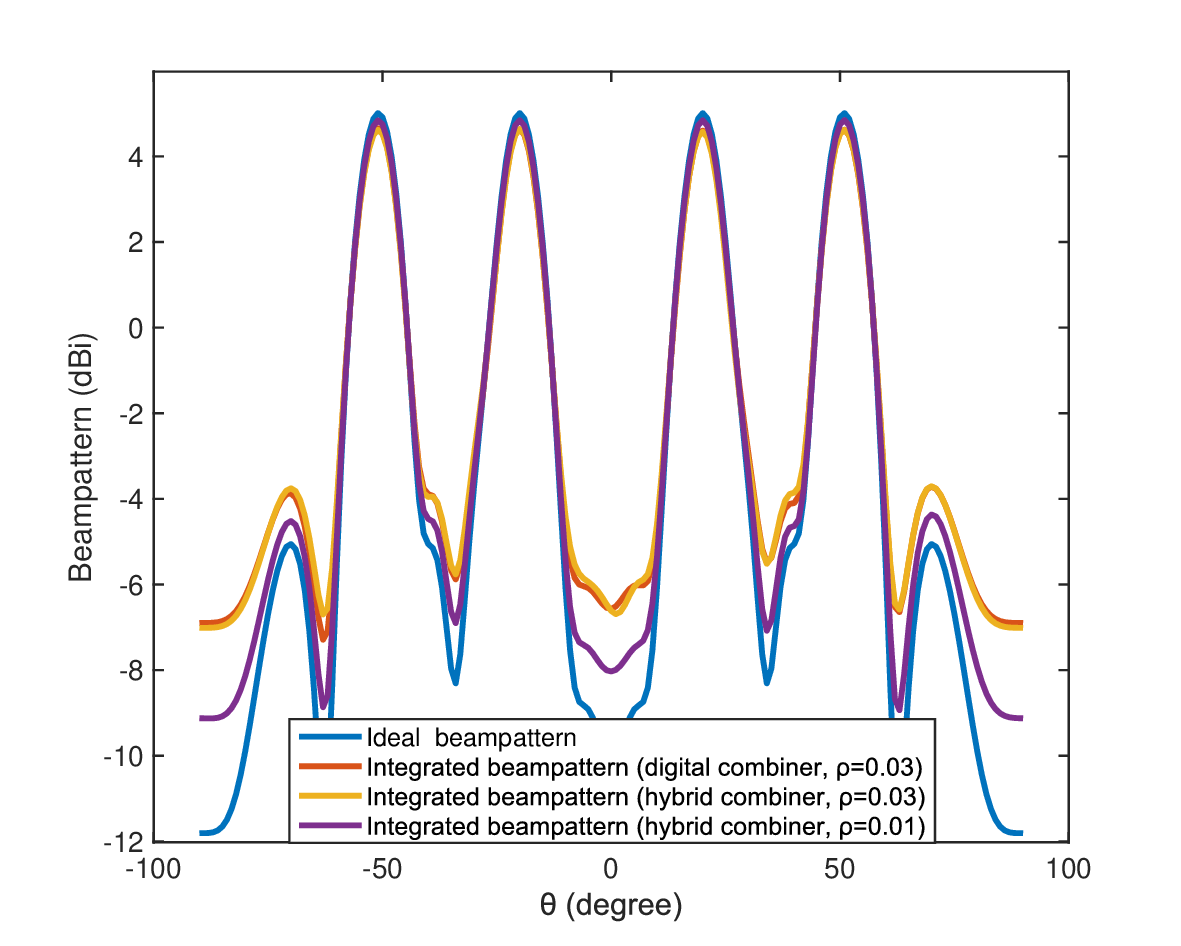}}
\caption{Integrated beampatterns of different schemes.}
\label{fig2}
\end{figure}

\begin{figure}[htbp]
\centerline{\includegraphics [width=3.0 in]{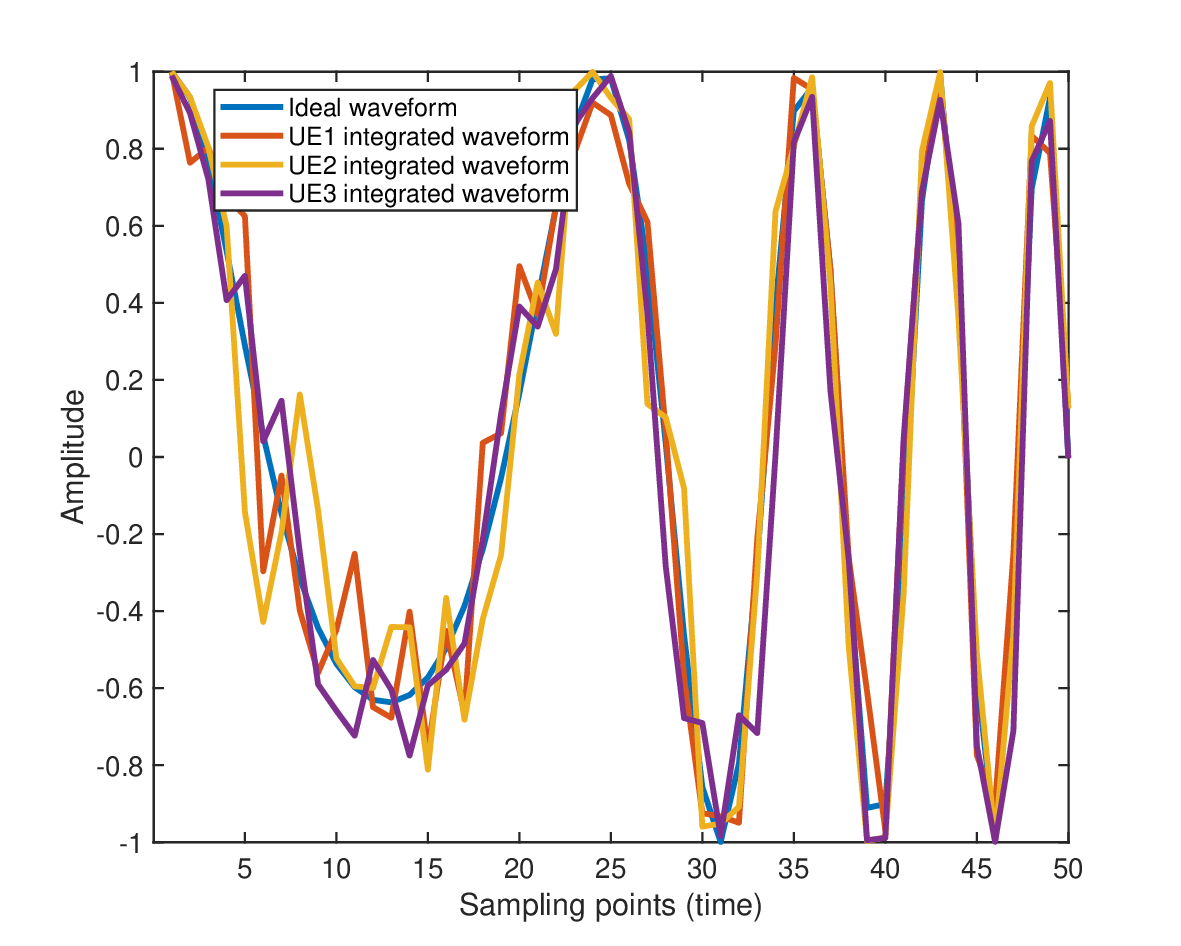}}
\caption{Integrated waveform.}
\label{fig3}
\end{figure}

In Fig. \ref{fig2}, we initially present the beampatterns derived from the proposed HBF design, as compared with several benchmarks. We can find that the hybrid combiner leads to nearly the same beampattern as that obtained from the digital combiner. This excellent match demonstrates that the proposed HBF scheme with the RCG algorithm can achieve satisfied performance with much lower complexity. As for the tradeoff setting $\rho$, it is noteworthy that the hybrid combiner can approximate a more similar beampattern to the ideal one by decreasing this parameter from $0.03$ to about $0.01$. We will further analyze the sensitivities of different performances to the parameter $\rho$ in next subsection. In addition, Fig. \ref{fig3} shows the designed waveform transmitted by one of the antennas of three UEs, and it aligns closely with the reference signal. 

\subsection{Convergence Performance and Pareto Bounds}

\begin{figure}[t]
\centerline{\includegraphics [width=3.0 in]{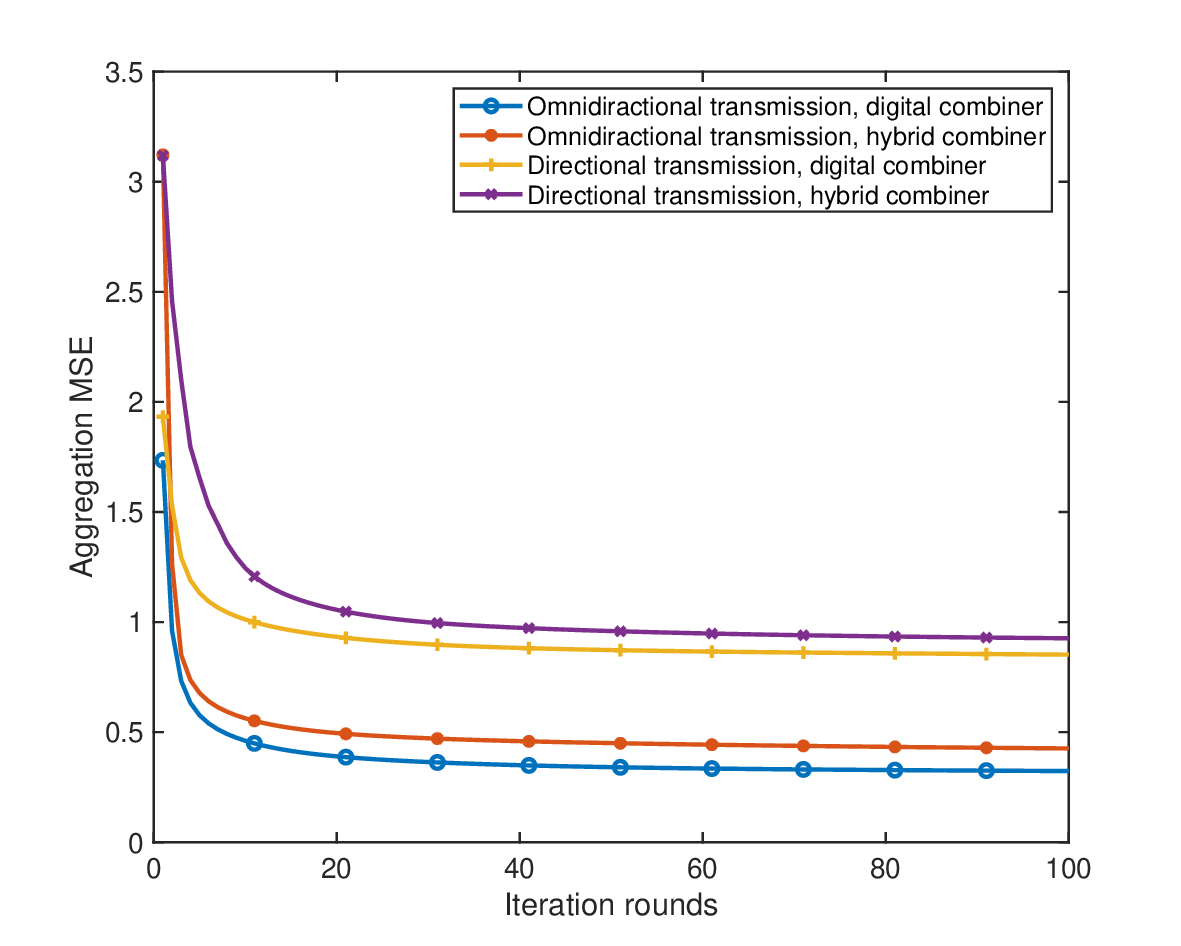}}
\caption{Convergence performance.}
\label{fig4}
\end{figure}

\begin{figure}[t]
\centerline{\includegraphics [width=3.0 in]{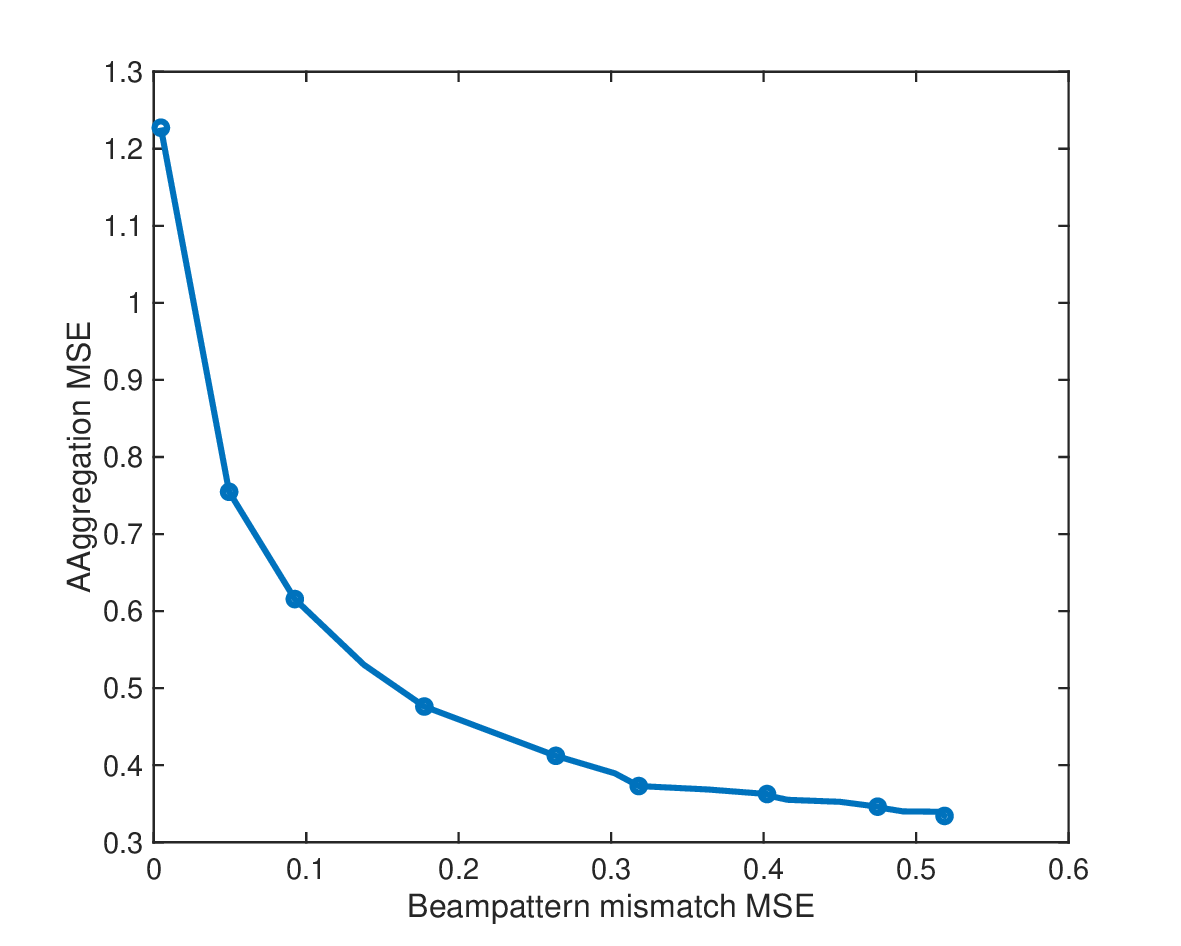}}
\caption{Pareto bounds.}
\label{fig5}
\end{figure}

Fig. \ref{fig4} illustrates the convergence performance of the proposed AO framework. It is observed that both schemes can converge in a few iterations, while the HBF scheme has just a slightly slower convergence speed and slightly worse MSE performance, since it has fewer independent RF streams to aggregate information. This also aligns well with the conclusion in \cite{yu2016alternating} that the HBF can approach the performance
of the fully digital beamforming when $N_{rf}$ is comparable to $M$. Furthermore, the directional transmission has better performance than the omnidirectional ones. This is because the power of directional transmission is more concentrated, leading to higher SNR at the AP. As a byproduct of the weighting optimization problem, the Pareto optimal points in Fig. \ref{fig5} provide more insight into the achievable bounds of the over-the-air aggregation and other performance. The upper right corner of this graph is the feasible solution area, and the points that build up the Pareto frontier are obtained by changing the $\rho$ over $[0.01,1]$ with the gap $0.05$. For the points on the frontier, improving one kind of performance will inevitably affect another performance of them. 

\subsection{Over-the-air aggregation Performance of integrated signal}

\begin{figure}[t]
\centerline{\includegraphics [width=3.0 in]{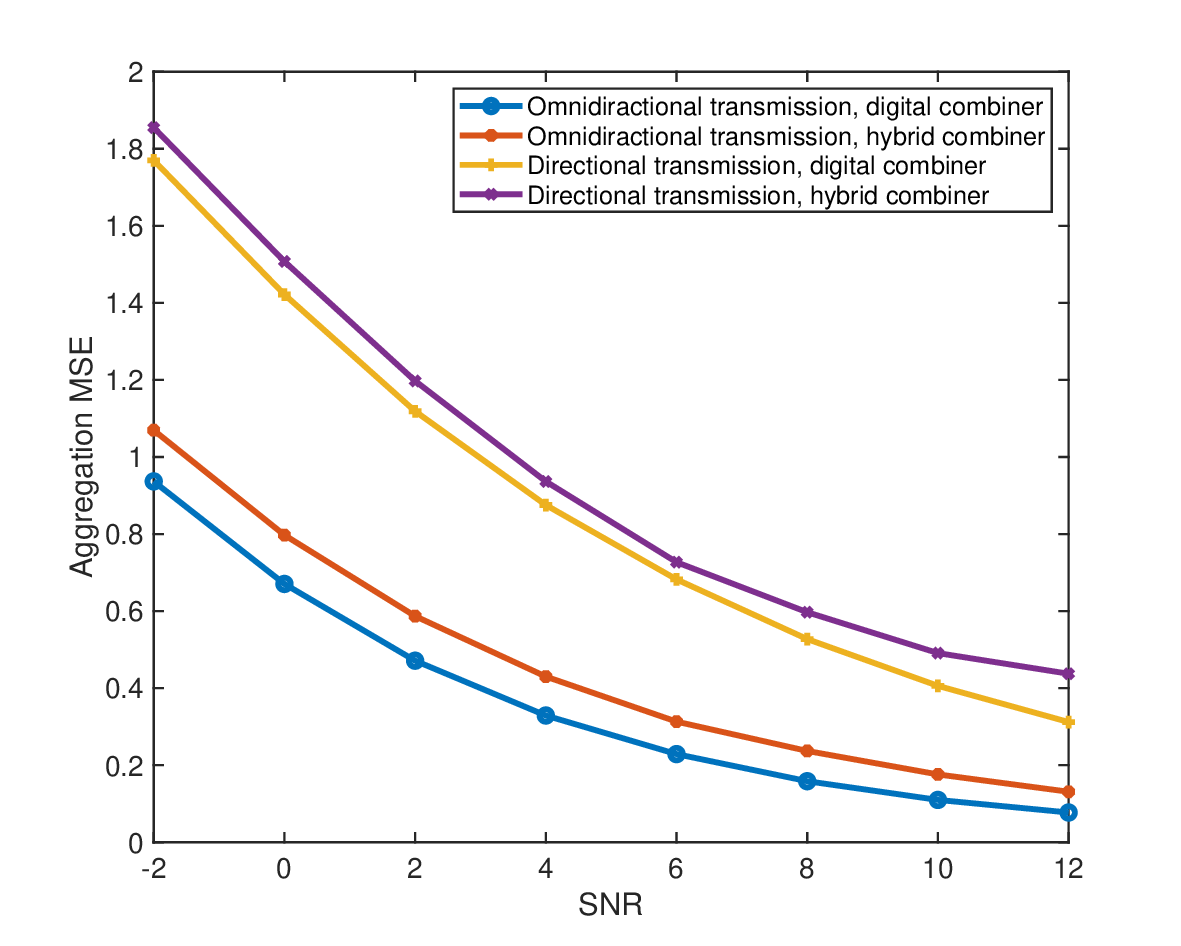}}
\caption{Normalized over-the-air aggregation MSE versus SNR.}
\label{fig6}
\end{figure}

\begin{figure}[t]
\centerline{\includegraphics [width=3.0 in]{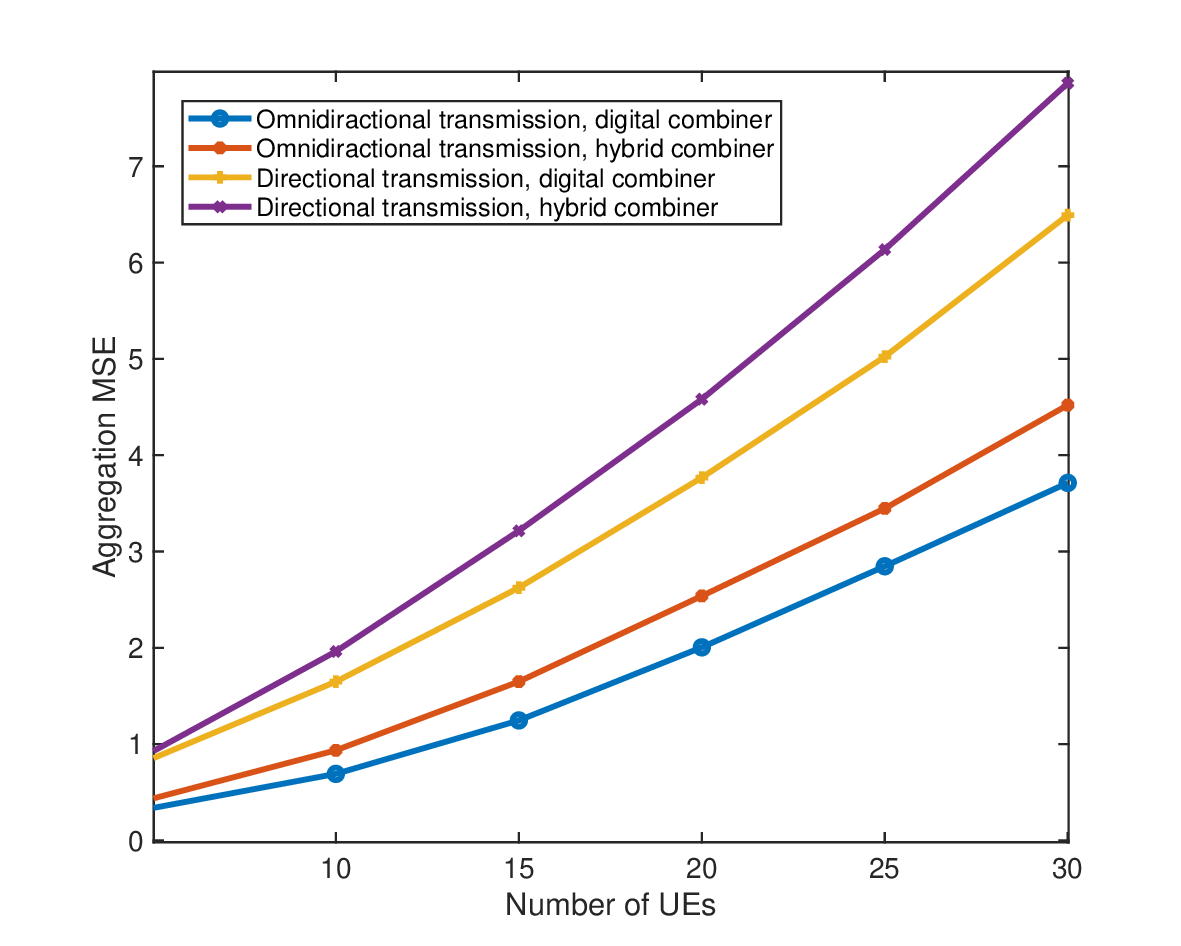}}
\caption{Normalized over-the-air aggregation MSE versus the number of UEs.}
\label{fig7}
\end{figure}

\begin{figure}[h]
\centerline{\includegraphics [width=3.0 in]{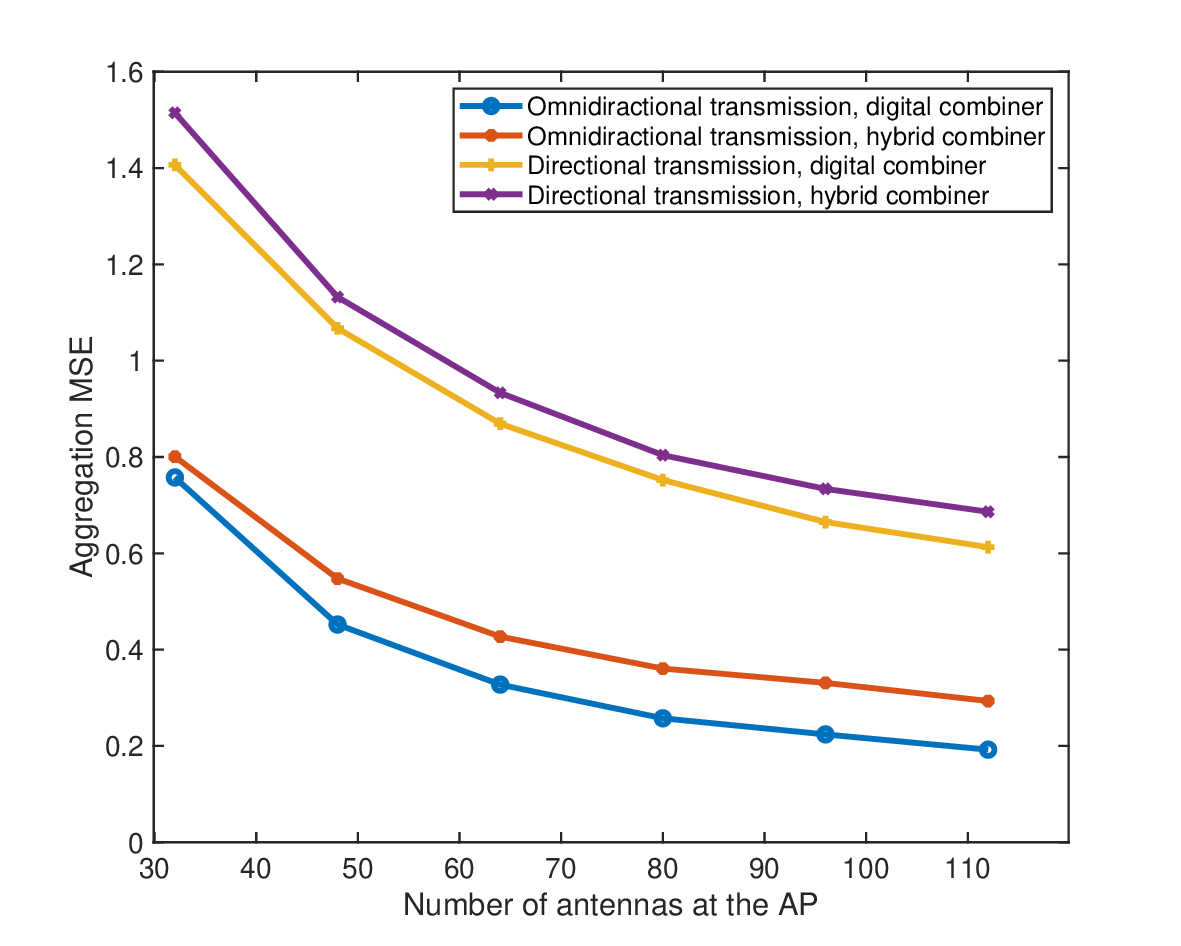}}
\caption{Normalized over-the-air aggregation MSE versus the number of antennas at the AP.}
\label{fig8}
\end{figure}

\begin{figure}[h]
\centerline{\includegraphics [width=3.0 in]{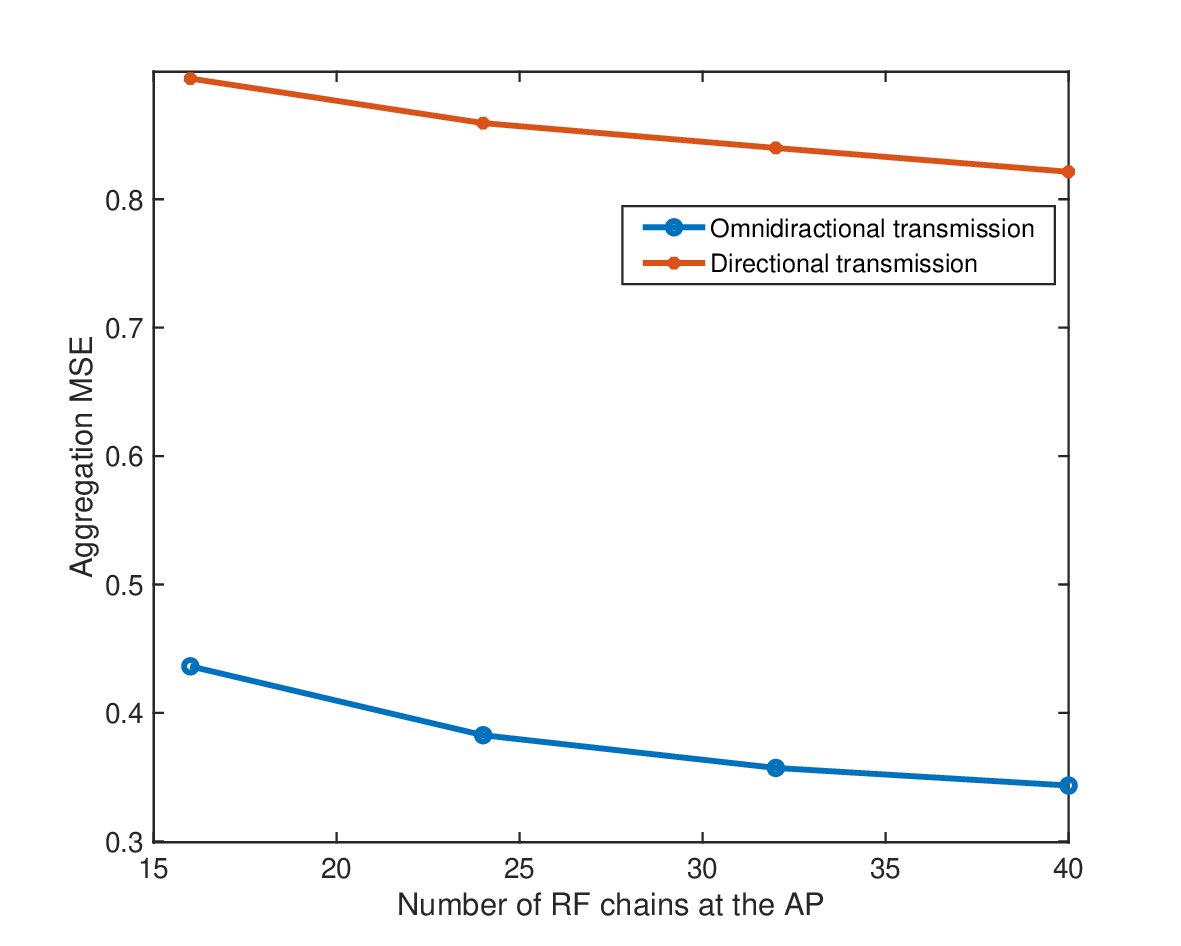}}
\caption{Normalized over-the-air aggregation MSE versus the number of RF chains at the AP.}
\label{fig9}
\end{figure}

In this subsection, we consider the relationship between over-the-air aggregation performance and some typical factors of integrated signal. It is depicted in Fig. \ref{fig6} that the MSE monotonically decreases with the SNR increases. In Fig. \ref{fig7}, we show the increasing number of UEs will result in higher normalized MSE. It becomes more challenging to appropriately design a common aggregation beamforming matrix to equalize the channels between AP and all the UEs. In this case, digital combiner shows stronger robustness than hybrid combiner. Fig.\ref{fig8} illustrates the curves of the aggregation MSE versus the number of antennas at each UE. We can observe that normalized MSE increases with the increasing number of antennas since more antennas can provide an extra degree of freedom (DoF) to optimize the beamforming performance and thus more significant array gain can be achieved. Fig. \ref{fig9} further compares the impact of different numbers of RF chains in HBF. Nevertheless, the increasing RF chains bring a little benefit to the HBF scheme, because the main limitation is no longer the number of RF chains when it already larger than $M$. In this case, there is no need to deploy more RF chains to increase cost. 

It can be concluded that, in both omnidirectional transmission and directional transmission cases, the proposed receive HBF with less processing overhead can achieve comparable performance with the digital scheme.

\section{Conclusion}
In this paper, we focus on the integrated HBF design and waveform design to enable simultaneous three kinds of tasks. As for the HBF design, a joint transceiver beamforming design based on the AO approach is proposed using SCA and RCG algorithms. In the waveform design, the reference signal is selected as the benchmark, and an SDR-based method is developed to optimize the waveform under CMC and SC. Numerical results have demonstrated that the above designs can achieve satisfactory performance while guarantee cost-effective and low complexity algorithm.    

\bibliographystyle{IEEEtran}
\bibliography{IEEEabrv, references}{}

\vfill

\end{document}